\documentclass{preprint}

\usepackage[T1]{fontenc}
\usepackage{times}

\usepackage{microtype,xurl}
\usepackage{comment}

\newenvironment{proofof}[1]{\begin{proof}\textbf{of {#1}}}{\end{proof}}
\usepackage{thm-restate}

\allowdisplaybreaks[2]

\newcommand{\E}{\mathbb{E}}
\newcommand{\ind}{\mathbf{1}}
\newcommand{\OPT}{\operatorname{OPT}}
\newcommand{\pmax}{p^{\max}}
\newcommand{\pmin}{p^{\min}}
\newcommand{\surrA}{A^{\circ}}

\title{Regret Bounds for Market Making with Persistent Orders}
\hypersetup{pdftitle={Regret Bounds for Market Making with Persistent Orders},pdfauthor={},urlcolor=blue}
\author{
    \Name{Nicol\`o {Cesa-Bianchi}} \Email{nicolo.cesa-bianchi@unimi.it}\\
    \addr{Universit\`a degli Studi di Milano, Italy}
    \AND%
    \Name{Tommaso Cesari} \Email{tcesari@uottowa.ca}\\
    \addr{University of Ottawa, Ottawa, Canada}
    \AND%
    \Name{Roberto Colomboni} \Email{roberto.colomboni@bristol.ac.uk}\\
    \addr{School of Mathematics, University of Bristol, Bristol, United Kingdom}
    \AND%
    \Name{Luigi Foscari} \Email{luigi.foscari@unimi.it}\\
    \addr{Universit\`a degli Studi di Milano, Italy}
    \AND%
    \Name{Vinayak Pathak} \Email{path.vinayak@gmail.com}\\
    \addr{Computational Rational Agents Laboratory (CORAL)}
}

\begin{document}
\maketitle
\begin{abstract}
We study adversarial market making with orders that remain active until execution. The maker operates on two markets: a primary market where orders are made and a secondary market where the resulting positions are hedged. In each round, the maker posts a quote on the primary market before observing both primary and secondary market prices. An order executes when the primary price reaches its quote, and the resulting position is immediately offset at the secondary price. When both price sequences are chosen by an adversary over $T$ rounds, we give a simple deterministic policy with regret of order $\log T$ against the best fixed quote (static comparator) and $\sqrt T$ against the best quote for each submission date (dynamic comparator), plus two terms measuring exposure to future primary and secondary prices. These rates remain optimal even when both exposure terms vanish and lower bounds show that the exposure terms are necessary.
Our policy naturally extends to simultaneous bids and asks, where we show that static and dynamic regret are both characterized (up to constant factors) by the exposures terms (where the primary price exposure $V_T^{\mathrm{two}}$ now accounts for both bids and asks) and a term of order $\sqrt{V_T^{\mathrm{two}}\log T}$. Our policy is simple, efficient and fully parameterless.
We extend this setting by demonstrating how natural assumptions on the price processes lead to improved guarantees without changing the policy.
\end{abstract}

\section{Introduction}
\label{sec:introduction}
A market maker quotes prices at which they are willing to buy and sell an asset. After a trade, they can offset their position in another (secondary) market. The difference between the quoted price and the price of this offsetting trade determines the maker's profit. We revisit the simple sequential model of \citet{marketmakingwithoutregret}, where the maker submits a bid and ask pair to an incoming trader in each round. Unlike that setting, where an unaccepted quote does not remain available to later trades, we study what happens when: (1) orders remain active until they trade, (2) orders cannot be canceled.
In our model the maker continues submitting orders even though the previous ones may still be outstanding. An ask therefore determines both a sale price and an execution date: a higher ask may earn more if it trades, but it may trade later.
Hence, even though every executed position is immediately offset, outstanding orders remain exposed to future prices, and many orders can become exposed to the same later change in the hedging cost.
Our model has two exogenous price sequences (one for the primary and one for the secondary market), both revealed after each decision. An outstanding ask executes the first time the primary price becomes at least as large as its quote, and the maker buys back the asset at that round's secondary price. The execution delay of outstanding asks is monotonic in their quote, as a lower quote cannot execute later than a higher one. Finally, our setting also includes bids, which are defined symmetrically to asks.

\paragraph{Our results.}
We study two notions of regret: static regret, where we compare the maker with the best fixed quote submitted on every round, and dynamic regret, where the comparator chooses the best quote separately on every round and knowing the entire price path. In both cases, the regret after $T$ rounds is controlled by two measures of price exposure plus a horizon-dependent term. The first measure is defined on the primary prices and counts the width of the intervals of quotes that wait and eventually execute. The second measure is defined on the secondary prices and records the possible change between the last observed hedging price and a future execution price. These measures can be small even when some orders wait a long time. The horizon-dependent term, instead, is $\Theta(\log T)$ for static regret and $\Theta\big(\sqrt{T}\big)$ for dynamic regret.
These bounds are obtained by the same policy using two different learning rates that depend neither on the horizon nor on the price exposures. 
We also prove lower bounds showing that this difference between static and dynamic regret is necessary, even when the primary price never increases and the secondary price is constant. Separate constructions also establish the need for both exposure terms. 
Our policy is simple, efficient, and naturally extends to bid-ask pairs.

When the learner can post both bids and asks, the horizon-dependent term vanishes and is now replaced by a primary exposure term on both sides which controls both upward and downward movements. While this does not mean that the two-sided problem is easier than the one-sided counterparts, we characterize both static and dynamic regret by the same rate, meaning that the adversary does not gain by adaptively choosing their quotes. We naturally extend the strategy for the ask-only case and show that the resulting policy is optimal for both types of regret simultaneously.

Further, in Appendix~\ref{app:structured} we investigate the impact of additional assumptions on the price processes and their impact on regret in the ask-only case. In particular, we study what happens when both processes orbit around fixed prices. Then we study the case in which the secondary price is equal to the primary, but with a one-round \textit{lag}~\citep{lo1990trading}. In both cases, the policy naturally ensures improved guarantees without knowledge of the assumptions.




\section{Related work}
\label{sec:related}

Adversarial formulations provide an alternative foundation for financial problems traditionally studied under stochastic price models. A seminal work in the same vein is by \citet{abernethy2013hedge}, who study option hedging as a sequential game against an adversary whose price changes satisfy a cumulative quadratic volatility budget and a bound on individual jumps. For convex, Lipschitz, eventually linear payoffs, they show that the minimax option price converges to the Black–Scholes price under suitable vanishing-jump conditions.
Their result demonstrates that a classical pricing benchmark can emerge without assuming a stochastic law for price evolution, motivating the study of financial decisions through guarantees expressed in terms of price-path properties.
We pursue this perspective for market making with persistent orders.
%

The model perhaps closest to ours is by \citet{marketmakingwithoutregret}. In both settings, a maker quotes a bid and an ask, and offsets each accepted trade in a secondary market. In their setting, the trader's private valuation determines acceptance and plays the role of our primary price, while the market price determines the offsetting cost and plays the role of our secondary price. They assume that an accepted quote trades and is immediately hedged, while an unaccepted quote expires. We study the opposite case in which an unfilled order remains available to later trades, cannot be canceled, and may never execute. Both models hedge immediately after execution.

The other difference between the two settings is the feedback. There, the maker observes the market price and whether the trader buys, sells, or does neither, but never observes the valuation. Their algorithms learn profitable quotes from this partial feedback by combining methods for first-price auctions and dynamic pricing, with an additional step that keeps the bid below the ask. Here both prices are revealed after every decision, even when nothing trades. This also corresponds to a form of partial observability, as the maker can compute the counterfactual reward of past quotes that would have executed at the current primary price.
While the main difficulty in the setting of \citet{marketmakingwithoutregret} arises from the partial feedback, here the difficulty is tied to the exposure to the unknown future prices.
\citet{maranOnlineMarketMaking2026} then study an intermediate feedback model in which the trader's valuation is revealed only when no trade occurs.

\citet{abernethy2013} on the other hand study adversarial market making with observed prices, competing with a class of spread-based strategies. Their learner must account for the inventory accumulated by each strategy when switching between them. Our immediate hedge removes this inventory state, while persistence creates commitments through outstanding orders.
Classical models instead explain spreads through adverse selection \citep{glosten1985} or inventory risk \citep{avellaneda2008}. Here future execution prices are the source of exposure despite immediate hedging.

Our work is related to the extensive literature on online learning with delayed feedback, in particular when delays are allowed to depend on the action, \citet{vanderhoeven2022} study this case in both the experts and bandit settings. More generally, \citet{joulani2013} give reductions from delayed to standard online learning, \citet{quanrud2015delays} analyze gradient and perturbed-leader methods with adversarial delays, and \citet{zimmert2020} obtain optimal regret bounds for adversarial bandits with arbitrary delays. These works show how delayed observations affect learning, often through bounds involving cumulative delay or outstanding feedback. \citet{vernade2017conversions} study stochastic bandits motivated by online advertising, where an ad may lead to a purchase realized only after a random, possibly infinite, delay. In our model, the execution time depends on both the posted quote and the primary price path, and determines which secondary price enters the reward, Thus, delays affect the payoff and the time feedback is available.

The idea of a delayed reward connects to the notion of losses with memory (e.g., switching costs), where the instantaneous loss depends not only on the current action, but also on the actions in a window on the past. \citet{DekelTA12,cesa2013online} show that a memory buffer of size $o(T^{1/3})$ is necessary and sufficient for sublinear regret, even with full feedback, when no other restrictions are imposed on the loss functions. In contrast, our instantaneous payoff is determined by a specific function of the entire past history parameterized by the adversarial choice of the primary and secondary price paths. As a result, our regret can be sublinear, depending on how fast the resulting price exposures grow with time.

Finally, our notion of primary market exposure is related to the notion of upcrossing studied by \citet{koolenBuyLowSell2014}. However, rather than bounding the investor's regret, their goal is to derive an exact characterization of how much wealth an investor can guarantee from the occurrence of price upcrossings, without knowing when or where those upcrossings will occur.


\section{Setting}
\label{sec:model}

Fix a horizon $T\ge2$.
An adversary chooses primary prices $p_1,\ldots,p_T\in[0,1]$ and secondary prices $m_1,\ldots,m_T\in[0,1]$. If the learner is randomized, then the adversary may be nonoblivious, i.e., adapt to the learner's past actions.
For simplicity, we initially restrict our analysis to ask quotes only, and later extend it to ask-bid pairs. On round $t$, the maker submits one unit at a price $A_t\in[0,1]$, using past observations and possibly internal randomness. The pair $(p_t,m_t)$ is then revealed. Every outstanding ask $A_s$, for $s \le t$, executes if $A_s\le p_t$. The maker then sells one unit at $A_s$ and immediately buys one unit in the secondary market at $m_t$. Each order executes at most once and orders cannot be canceled.

%
Both the primary and secondary price are revealed at the end of the round even when no trading happens, therefore the maker observes more than bandit feedback, although the final reward of a submitted order can remain unknown, as an order still outstanding after the time horizon receives zero reward. Bids, and the constraint relating them to asks, are introduced in Section~\ref{sec:two-sided}.

For an ask $a$ submitted at $t$, its execution time is
$
 \tau_t(a):=\min\{s\in\{t,\ldots,T\}:p_s\ge a\}
$
where the minimum of the empty set is $\infty$.
When the ask executes, its execution cost is $c_t(a):=m_{\tau_t(a)}$, while its reward is
\begin{equation}
 r_t(a):=
 \begin{cases}
 a-c_t(a),&\tau_t(a)\le T,\\
 0,&\tau_t(a)=\infty.
 \end{cases}
 \label{eq:reward}
\end{equation}
The subscript records the submission date, not the execution date; in particular, several asks may be traded on the same round. The boundary ask $a=1$ always has nonnegative reward because either it does not trade or it earns $1-c_t(1)\ge0$.

Let $\pmax_t:=\max_{t\le s\le T}p_s$ be the highest \textit{remaining} primary price, using this notation, the reward~\eqref{eq:reward} can be equivalently defined by $r_t(a):=(a-c_t(a))\ind\{a\le \pmax_t\}$. An ask submitted at $t$ executes by $T$ if $a\le \pmax_t$, note that the sequence $(\pmax_t)_t$ is nonincreasing. Importantly, if $a\le a'$, then $\tau_t(a)\le\tau_t(a')$, meaning that a lower ask cannot wait longer than a higher one submitted on the same round.

\paragraph{Benchmarks.}
We define regret with respect to two different comparators.
The \textit{fixed} benchmark submits the same ask on every round, while the \textit{dynamic} benchmark may choose a different ask for each submission date, and both obey the same execution and horizon rules as the maker. Denote the optimal comparators and the corresponding regrets as
\begin{align}
 \OPT_T&:=\max_{a\in[0,1]}\sum_{t=1}^T r_t(a)~,
 &\OPT_T^{\mathrm{dyn}}&:=\sum_{t=1}^T\max_{a\in[0,1]}r_t(a)~,
 \label{eq:benchmarks}\\
 R_T&:=\E\!\left[\OPT_T-\sum_{t=1}^T r_t(A_t)\right],
 &R_T^{\mathrm{dyn}}&:=\E\!\left[\OPT_T^{\mathrm{dyn}}
       -\sum_{t=1}^T r_t(A_t)\right].
 \label{eq:regrets}
\end{align}
Expectations are over the (possible) maker's randomness.
The safe boundary ask $a = 1$ ensures that both benchmarks are nonnegative, and $R_T\le R_T^{\mathrm{dyn}}$.
The maxima in~\eqref{eq:benchmarks} exist. To see this, round any ask up to the smallest point of $\{p_1,\ldots,p_T,1\}$ that is at least that ask. For every submission date, this preserves execution, while maximizing the sale price. Thus a finite set contains an optimizer for either benchmark.

\paragraph{Market exposure.}
%
Define the primary and secondary cumulative exposures as
\begin{equation}
 V_T:=\sum_{t=1}^{T-1}(\pmax_t-p_t),\qquad
 w_t:=\max_{t<s\le T}|m_s-m_t|,\qquad
 W_T:=\sum_{t=1}^{T-1}w_t.
 \label{eq:exposures}
\end{equation}
The asks submitted at $t$ that execute later live in the interval $(p_t,\pmax_t]$ of length $\pmax_t-p_t$, thus $V_T$ measures the total width of delayed execution opportunities, rather than the number of rounds spent waiting. The quantity $w_t$ bounds the change from $m_t$, which is the last observed secondary price when choosing $A_{t+1}$, to any possible later execution cost. Both exposures are properties of the price paths and are unknown to the maker. Intuitively, low exposure terms mean more predictable price sequences, which facilitate trading.
For example, a nonincreasing primary path and a constant secondary path have $V_T=W_T=0$.
A single secondary-price change of size $\delta$ at round $k$, followed by a constant price, gives $W_T=(k-1)\delta$. The same change therefore has a larger exposure when it occurs later. More generally,
\begin{equation}
 W_T\le\sum_{k=2}^T(k-1)|m_k-m_{k-1}|
 \label{eq:weighted-variation}
\end{equation}
by applying the triangle inequality to each future movement. In the following, we use $H_t:=\sum_{s=1}^t1/s = \Theta(\log t)$ for the harmonic sum and $(x)_+:=\max\{x,0\}$ for the positive part.

\section{Static regret}
\label{sec:fixed}

We now provide a policy to compete against the best fixed quote achieving $\OPT_T$.
Initially the policy quotes $A_1 = 1$. Then, for any $t \ge 2$ and for any sequence $\big(\lambda_t\big)_{t\ge 2}$ of numbers $\lambda_t\in [0, 1]$, the policy sets
\begin{equation}
\label{eq:policy}
    A_t
=
    m_{t-1} + {(1-\lambda_t)} (p_{t-1} - m_{t-1})_+~.
\end{equation}
The policy uses the last observed primary and secondary prices as proxies for quote and cost. If these proxies suggest a positive profit ($p_{t-1} > m_{t-1}$), then the policy quotes a price $A_t$ in between the two prices, otherwise the policy sets $A_t = m_{t-1}$.
A larger $\lambda_t$ sacrifices more margin when an order fills, but makes it harder for a small threshold decrease to prevent execution. For static regret, where we compete against the best fixed quote, we use $\lambda_t  = 1/t$, so that $A_t$ converges to ${\max\{p_{t-1},m_{t-1}\}}$ as $t$ grows.

Note that this policy requires constant time and memory to choose each new quote and does not need the horizon or either exposure as input.
%
\begin{theorem}
\label{thm:fixed}
For every pair of price paths, policy~\eqref{eq:policy} with $\lambda_t = 1/t$ satisfies
\[ R_T\le H_T+V_T+2W_T \,. \]
\end{theorem}
In particular, $V_T+W_T=o(T)$ is sufficient for sublinear regret. When the primary price never increases and the secondary price is constant, the bound is $H_T = \Theta(\log T)$, even if the primary price falls by a constant amount.

We prove the theorem using the last observed secondary price $m_{t-1}$ as a proxy for the future execution costs $c_t(\cdot)$. Define the surrogate rewards
\begin{equation}
    r_1'(a):=(a-1)\ind\{a\le \pmax_1\}~,\qquad
    r_t'(a):=(a-m_{t-1})\ind\{a\le \pmax_t\}\quad(t\ge2)~.
\label{eq:surrogate}
\end{equation}
Note that the condition determining execution remains unchanged. The fixed surrogate benchmark is $\OPT'_T:=\max_a\sum_t r_t'(a)$. Let $C_t:=1+\sum_{s=1}^{t-1}m_s$ be the cumulative reference cost. Because $\pmax_t$ is nonincreasing, a fixed ask $a$ with $\pmax_1 \ge a$ executes precisely in a prefix $[t]$ of the $T$ rounds for some $t \ge 1$. The surrogate profit obtained by playing $a$ every round is $ta-C_t\le t\pmax_t-C_t$. Hence we upper bound the surrogate optimum by considering the most profitable prefix
\begin{equation}
 \OPT'_T\le
 \max\left\{0,\max_{1\le t\le T}(t\pmax_t-C_t)\right\}.
 \label{eq:prefix}
\end{equation}
The regret on the surrogate rewards is best understood using a \emph{clairvoyant} policy: this is the policy~\eqref{eq:policy} run on the surrogate rewards with $\pmax_{t-1}$ instead of $p_{t-1}$. The policy sets $\surrA_1=1$ and, for all $t \ge 2$ and any sequence $\lambda_t \in [0, 1]$,
\begin{equation}
\label{eq:auxiliary-fixed}
    \surrA_t
=
    m_{t-1} + {(1-\lambda_t)} (\pmax_{t-1} - m_{t-1})_+~,
\qquad
    Q_t^\circ:=\sum_{s=1}^t r_s'(\surrA_s)~.
\end{equation}

From~\eqref{eq:surrogate} one can see that surrogate rewards ${r_t'}(\surrA_t)$ are always nonnegative.

Let $R_T(A)$ be the regret of policy~\eqref{eq:policy}, $R_T'$ the surrogate regret, and ${R_T'}(\surrA)$ the surrogate regret of the clairvoyant policy~\eqref{eq:auxiliary-fixed}. Theorem~\ref{thm:fixed} is an immediate consequence of the following results:
\begin{align*}
    R_T(A) &\le R_T'(A) + 1 + 2W_T~,   \tag{Lemma~\ref{lem:execution-cost}}
\\
    R_T'(A) &\le R_T'(\surrA) + V_T~,  \tag{Lemma~\ref{lem:fixed-surrogate}}
\\
     R_T'(\surrA) &\le H_T - 1~.   \tag{Lemma~\ref{lem:potential}}
\end{align*}
\begin{lemma}
\label{lem:execution-cost}
For any choice of $\big(\lambda_t\big)_{t\ge 2}$, policy~\eqref{eq:policy} satisfies $R_T(A) \le R_T'(A) + 1 + 2W_T$.
\end{lemma}
\begin{proof}
For every ask and $t\ge2$,
\begin{equation}
    |r_t(a) - r_t'(a)|
=
    |m_{t-1} - m_{\tau_t(a)}|\ind\{a\le \pmax_t\}
\le
    w_{t-1}~.
 \label{eq:transfer}
\end{equation}
At $t=1$, $r_1(a)\le r_1'(a)+1$, while the learner has $r_1(1)\ge r_1'(1)=0$. Thus
\[
    \OPT_T-\sum_{t=1}^T r_t(A_t)
\le
    \OPT_T'-\sum_{t=1}^T r_t'(A_t) + 1 + 2W_T~,
\]
concluding the proof.
\end{proof}
Next, we prove $R_T'(A) \le R_T'(\surrA) + V_T$ via the following lemma.
\begin{lemma}
\label{lem:fixed-surrogate}
For any choice of $\big(\lambda_t\big)_{t\ge 2}$, policies~\eqref{eq:policy} and~\eqref{eq:auxiliary-fixed} satisfy
\[
    \sum_{t=1}^T r_t'(A_t)
\ge
    \sum_{t=1}^T r_t'(\surrA_t) - V_T~.
\]
\end{lemma}
\begin{proof}
By comparing~\eqref{eq:policy} and~\eqref{eq:auxiliary-fixed} one can see that for all $t\ge2$, the actual quote $A_t$ lies between $m_{t-1}$ and $\surrA_t$. Lowering an ask within this interval preserves a nonnegative surrogate margin and can only make execution more likely. If both quotes $A_t,\surrA_t$ execute, the loss is their price difference $\surrA_t-A_t$; if $\surrA_t$ does not execute, then $r_t'(\surrA_t)=0$. Therefore
\begin{align*}
    r_t'(A_t)
\ge
    r_t'(\surrA_t) -(\surrA_t-A_t)
\ge
    r_t'(\surrA_t) - (1-\lambda_t)(\pmax_{t-1}-p_{t-1})~.
\end{align*}
Summing over $t=1,\ldots,T$ gives
\begin{align*}
    \sum_{t=1}^T r_t'(A_t) - \sum_{t=1}^T r_t'(\surrA_t)
\ge
     - \sum_{t=1}^{T-1}(1-{\lambda_{t+1}})(\pmax_t-p_t)
\ge
    - \sum_{t=1}^{T-1}(\pmax_t-p_t)
=
    {-V_T}~,
\end{align*}
concluding the proof.
\end{proof}

Finally, we prove $R_T'(\surrA) \le H_T - 1$. Recall that the quantity $t\pmax_t-C_t$ in~\eqref{eq:prefix} bounds the surrogate profit of a fixed ask that executes for the first $t$ submission dates. 
%
\begin{lemma}
\label{lem:potential}
For every $t \in [T]$, if the auxiliary policy~\eqref{eq:auxiliary-fixed} is run with $\lambda_t = 1/t$, then it satisfies $Q_t^\circ\ge t\pmax_t-C_t-H_t+1$. Consequently, $R_T'(A^\circ) \le H_T-1$.
\end{lemma}
The proof is given in Appendix~\ref{app:potential}.

The primary price exposure $V_T$ admits a natural interpretation in terms of \textit{upcrossings}. A price path upcrosses an interval $[a,b]$ if it visits a price at most $a$ before subsequently reaching a price at least $b$. Hence, $V_T$ is the total mass of all upcrossings $[p_t,\pmax_t]$ of the primary price path. \citet{koolenBuyLowSell2014} use upcrossings to characterize exactly which wealth guarantees can be achieved by an online trading strategy.
While their guarantees only depend on the existence and magnitude of price upcrossings, $V_T$ also captures their multiplicity across outstanding orders. In particular, a single late price increase may generate exposure proportional to $T$ by executing many previously submitted orders.


\section{Dynamic regret}
\label{sec:dynamic}
In this section we extend the previous analysis to dynamic regret~\eqref{eq:regrets}. Surprisingly, the optimal policy~\eqref{eq:policy} for static regret is also optimal for dynamic regret provided the tuning $\lambda_t = 1/t$ is changed to $\lambda_t = 1/\sqrt{t-1}$. Indeed, the choice $\lambda_t = 1/t$ in~\eqref{eq:policy} can lead to linear dynamic regret, even when both exposures vanish (Appendix~\ref{app:separation} gives an example).
Next, we state and prove our main result for this section.
\begin{theorem}
\label{thm:dynamic}
For every pair of price paths, policy~\eqref{eq:policy} with $\lambda_t = 1/\sqrt{t-1}$ satisfies
\[
    R_T^{\mathrm{dyn}}
\le
    1+3\sqrt{T-1}+V_T+2W_T \,.
\]
\end{theorem}
\begin{proof}
Similarly to the proof of Theorem~\ref{thm:fixed}, we use surrogate rewards~\eqref{eq:surrogate} and the clairvoyant strategy~\eqref{eq:auxiliary-fixed}. We start by computing the optimal dynamic strategy for the surrogate reward. For any $t\ge2$, if $\pmax_t<m_{t-1}$, the safe boundary ask $1$ attains zero reward. Otherwise the ask $\pmax_t$ attains $(\pmax_t-m_{t-1})$. Hence, the surrogate reward of the optimal dynamic strategy at each $t \ge 2$ is $(\pmax_t-m_{t-1})_+$.

Now replace $p_{t-1}$ in~\eqref{eq:policy} by $\pmax_{t-1}$ to obtain the clairvoyant quote $\surrA_t$. By Lemma~\ref{lem:fixed-surrogate},
\begin{equation}
\label{eq:trivial-adaptation}
    \sum_{t=1}^T r_t'(A_t)
\ge
    \sum_{t=1}^T r_t'(\surrA_t) - V_T~. 
\end{equation}
Hence, the additional surrogate dynamic regret from using the actual quotes $A_t$ is at most $V_T$.
Fix $t\ge2$, and abbreviate $x=(\pmax_{t-1}-m_{t-1})_+$ and $y=(\pmax_t-m_{t-1})_+$. We proceed by case analysis.
\begin{enumerate}
    \item If $\surrA_t$ executes, Then the surrogate reward is $(1-\lambda_t)x$. Since $y\le x\le1$, the instantaneous dynamic regret is $y-(1-\lambda_t)x \le x - (1-\lambda_t)x \le \lambda_t$.
    \item If $\surrA_t$ does not execute, when $y=0$, the instantaneous regret is zero, otherwise, $y > 0$ and
\[
    (1-\lambda_t)x = (1-\lambda_t)(\pmax_{t-1}-m_{t-1})_+ > \pmax_t - m_{t-1} = y > 0
\]
so, in this case, 
$
    \pmax_{t-1}-\pmax_t=x-y>\lambda_t x\ge\lambda_t y
$
and the instantaneous regret is therefore at most ${y} < (\pmax_{t-1}-\pmax_t)/\lambda_t$.
\end{enumerate}
Combining the two results, we have that
\begin{equation}  
\label{eq:combining}
    \max_{a \in [0,1]} r_t'(a) - r_t'(\surrA_t) \le \lambda_t + (\pmax_{t-1}-\pmax_t)/\lambda_t
\end{equation}
Therefore,
\begin{align*}
    R_T^{\mathrm{dyn}}
&\le
    {1 +} \sum_{t=1}^T \max_{a \in [0,1]} r_t'(a) - \sum_{t=1}^T r_t'(A_t) + 2W_T \tag{{proof of} Lemma~\ref{lem:execution-cost}}
\\&\le
    {1 +} \sum_{t=1}^T \max_{a \in [0,1]} r_t'(a) - \sum_{t=1}^T r_t'(\surrA_t) + V_T + 2W_T \tag{using~\eqref{eq:trivial-adaptation}}
\\&\le
    1 + \sum_{t=2}^T \left(\lambda_t + \frac{\pmax_{t-1}-\pmax_t}{\lambda_t}\right) + V_T + 2W_T \tag{using~\eqref{eq:combining}}
\\&\le
    1 + 2\sqrt{T-1} + \sqrt{T-1} + V_T + 2W_T
\end{align*}
where we used $\sum_{t=2}^T\lambda_t \le 2\sqrt{T-1}$ and
\begin{align*}
    \sum_{t=2}^T \big(\pmax_{t-1}-\pmax_t\big)\sqrt{t-1}
&=
    \pmax_1 - \pmax_{{T}}\sqrt{T-1} + \sum_{t=2}^{{T-1}} \pmax_t\left(\sqrt{t} - \sqrt{t-1}\right)
\\ &\le
    \pmax_1 - \pmax_{{T}}\sqrt{T-1} + \pmax_1\sqrt{T-1} - \pmax_1
\\ &\le
    \sqrt{T-1}~.
\end{align*}
This concludes the proof.
\end{proof}

%

\section{Lower bounds}
\label{sec:lower}
We now show why changing the comparator changes the baseline rate from an harmonic to a square-root term, and why both exposure terms are needed. Every lower bound applies to randomized policies and can be realized by a deterministic oblivious price path.

To start, we show that the rates provided for static and dynamic regret are tight in $T$, showing that the dynamic problem is significantly harder.
\begin{theorem}
\label{thm:baseline-lower}
For every policy and for every $T\ge2$, there is a pair of price paths such that $R_T\ge\frac{3}{16}(H_T-1)$, and also $R_T^{\mathrm{dyn}}\ge\frac{3}{16}\sqrt{T-1}$ for a possibly different pair of paths. In both cases, the price paths are such that $V_T=W_T=0$.
\end{theorem}
\begin{proof}
Set $m_t=0$ for all $t$. The instance uses a sequence of values $\delta_t\in(0,1)$ which are specified below, these values define a random primary price process as
\[
    p_1 = 1,
    \qquad
    p_t = \begin{cases}
        (1 - \delta_t) p_{t-1} & \text{w.p. } \delta_t \\
        p_{t-1} & \text{w.p. } 1 - \delta_t
    \end{cases}
    \quad \text{for } t \ge 2 \, .
\]
Sample this sequence before play independently for each round. Every path is nonincreasing, so missed asks never execute later and $V_T=W_T=0$.

Given the full past with $p_{t-1}=x$, in expectation any ask earns at most $(1-\delta_t)x$, because lower quotes earn no more even if they trade, quotes in $((1-\delta_t)x,x]$ trade with probability $1-\delta_t$, and higher quotes never trade. The same holds once we randomize the quote, write $\mu_t:=\E[p_t]$ and set $\mu_0=\mu_1=1$ and $\delta_1=0$. Then $\mu_t=(1-\delta_t^2)\mu_{t-1}$, and the learner's expected total reward is at most $\sum_{t=1}^T(1-\delta_t)\mu_{t-1}$. We know consider the two cases.


\begin{itemize}
    \item  For dynamic regret, take $\delta_t=\delta:=1/(2\sqrt{T-1})$ for $t\ge2$. Because the secondary price is zero, the comparator quotes $p_t$ and earns it on every round, so its expected reward is $\mu_t$, while the learner earns at most $(1-\delta)\mu_{t-1}$. The expected regret of round $t$ is therefore
    \[
        \mu_t - (1-\delta)\mu_{t-1}
    =
        (1 - \delta^2)\mu_{t-1} - (1 - \delta)\mu_{t-1}
    =
        \delta(1-\delta)\mu_{t-1}~.
    \]
    Since $1-\delta\ge1/2$ and $\mu_{t-1}=(1-\delta^2)^{t-2}\ge1-(t-2)\delta^2\ge3/4$, summing over $t\ge2$ yields $\E[R_T^{\mathrm{dyn}}]\ge3\sqrt{T-1}/16$.

    \item For fixed regret, take $\delta_t=1/(2t)$ for $t\ge2$. Now $\mu_t=\prod_{s=2}^t(1-1/(4s^2))\ge1-\frac14\sum_{s=2}^{\infty}s^{-2}\ge3/4$, using $\sum_{s=2}^{\infty}s^{-2}\le\int_1^\infty x^{-2}\,dx=1$. The fixed quote $p_T$ earns $Tp_T$ through immediate execution on every round, so $\OPT_T\ge Tp_T$. The identity
    \[
     \sum_{t=1}^T\mu_{t-1}-T\mu_T
     =\sum_{t=1}^T t(\mu_{t-1}-\mu_t)
     =\sum_{t=1}^T t\delta_t^2\mu_{t-1}
    \]
    therefore gives
    \[
     \E[R_T]
     \ge T\mu_T-\sum_{t=1}^T(1-\delta_t)\mu_{t-1}
     = \sum_{t=1}^T \mu_{t-1}(\delta_t - t \delta_t^2)
     = \frac14\sum_{t=2}^T\frac{\mu_{t-1}}{t}
     \ge \frac3{16}(H_T-1)~.
    \]
\end{itemize}
This concludes the proof.
\end{proof}
Together with Theorems~\ref{thm:fixed} and~\ref{thm:dynamic}, these bounds give rates $\Theta(\log T)$ and $\Theta(\sqrt T)$ on the class $V_T=W_T=0$, for the two different comparators.

%
The next construction shows that the dependency on the primary exposure is tight, it works by postponing every profitable execution until the last round (maximum exposure). It is a posted-price problem with no chance to learn the terminal valuation before choosing any of the prices.
%
\begin{restatable}{theorem}{tcprimarylower}
\label{thm:primary-lower}
For every policy, $T\ge2$, and $v\in(0,T-1]$, there are deterministic price paths with $V_T\le v$, $W_T=0$, and $R_T\ge v/e$.
\end{restatable}

Take zero primary prices until the final round and zero secondary prices. Write $\beta=v/(T-1)$ and let the terminal price $X$ lie in $[\beta/e,\beta]$. The construction has $\mathbb P(X\ge a)=\beta/(ea)$ on this interval, so every ask there has expected reward $a\mathbb P(X\ge a)=\beta/e$ which is fixed for all quotes. Asks outside this interval earn no more. Meanwhile, $\E[X]=2\beta/e$, and the comparator quotes $X$ on every round. Thus the learner earns at most half the comparator's expected reward per order. See Appendix~\ref{app:primary-lower} for all the details.

Because secondary exposure can be large even when primary exposure is arbitrarily small, we next present a separate lower bound showing that, if the learner does not know whether the secondary price will rise or fall, they are forced to accumulate secondary exposure.

\begin{restatable}{theorem}{tcsecondarylower}
\label{thm:secondary-lower}
For every policy, $T\ge2$, and $w\in[0,(T-1)/4]$, there are deterministic price paths with $W_T=w$, $V_T= w / 4T$ and $R_T \ge 3w/8$.
Additionally, for $w>0$, the primary exposure can be made arbitrarily smaller while retaining $W_T=w$ and the regret bound.
\end{restatable}

Initially the prices are $3/4$ and $1/2$. On the final round the primary price rises slightly, while the secondary price either rises or falls by the same amount $w/(T-1)$. The fixed comparator delays trading when costs fall and trades early when they rise, while the maker must quote before knowing what is going to happen. Appendix~\ref{app:secondary-lower} gives the proof. Taking $w$ proportional to $T$ yields linear regret even as $V_T$ tends to zero.

The last two theorems also lower-bound dynamic regret, since $R_T^{\mathrm{dyn}}\ge R_T$.

Next we show that the previous three lower bounds can be easily combined to show that the dependency of policy~\eqref{eq:policy} on time and on both primary and secondary exposure terms is optimal.

\begin{theorem}
For every policy, $T\ge2$, $v\in(0,T-1]$, and $w\in[0,(T-1)/4]$, there is a deterministic price path with $V_T\le v$ and $W_T\le w$ such that
\[
    R_T \ge \max\left\{
        \frac{3}{16}(H_T-1),\, \frac{v}{e},\, \frac{3w}{8}
    \right\}
    = \Omega(\log T+v+w).
\]
Likewise, there is such a path with
\[
    R_T^{\mathrm{dyn}} \ge \max\left\{ \frac{3}{16}\sqrt{T-1},\, \frac{v}{e},\, \frac{3w}{8} \right\}
    = \Omega(\sqrt T+v+w).
\]
\end{theorem}

\begin{proof}
Apply Theorems~\ref{thm:baseline-lower}, \ref{thm:primary-lower}, and \ref{thm:secondary-lower}. Each construction can be chosen within the budgets $V_T\le v$ and $W_T\le w$, so the worst-case regret over this class is at least the maximum of the three lower bounds. Finally, the maximum of three nonnegative quantities is at least one third of their sum.
\end{proof}

\section{Simultaneous bids and asks}
\label{sec:two-sided}

We now focus on the case in which the maker submits both a bid $B_t$ and an ask $A_t$, with $0\le B_t\le A_t\le1$. A bid $b$ submitted at $t$ executes at $\tau_t^{\mathrm{bid}}(b):=\min\{s\in\{t,\ldots,T\}:p_s\le b\}$, with an empty minimum interpreted as $\infty$. Its reward $r_t^{\mathrm{bid}}(b)$ is $m_{\tau_t^{\mathrm{bid}}(b)}-b$ if executed and zero otherwise. We rename $r_t(a)$ to $r_t^{\mathrm{ask}}(a)$. The bid $0$ and the ask $1$ both have nonnegative rewards.

The orders created by the traders are independent and their execution is tied to the primary price. Because the outstanding quotes persist with time, it might be the case that bids and asks end up briefly overlapping, as the constraint $B_t\le A_t$ applies only to new pairs, ignoring previously posted quotes. However, at the end of any given round, all outstanding bids lie below the primary price and all outstanding asks lie above the primary price, thus a single change in the primary price triggers either a bid or an ask, not both. This makes the coupling of the two problems simple, allowing us to run independently one policy for the asks and another one for the bids.

Define the fixed and dynamic two-sided benchmarks by
\begin{align*}
\OPT_T^{\mathrm{two}}
&:=\max_{0\le b\le a\le1}\sum_{t=1}^T
\bigl(r_t^{\mathrm{bid}}(b)+r_t^{\mathrm{ask}}(a)\bigr)~,
&\OPT_T^{\mathrm{dyn,two}}
&:=\sum_{t=1}^T\max_{0\le b\le a\le1}
\bigl(r_t^{\mathrm{bid}}(b)+r_t^{\mathrm{ask}}(a)\bigr)~.
\end{align*}
Subtracting the learner's expected total reward defines $R_T^{\mathrm{two}}$ and $R_T^{\mathrm{dyn,two}}$, respectively. Maxima exist by rounding asks up and bids down to $\{0,1,p_1,\ldots,p_T\}$, preserving feasibility and execution dates. Let $p_t^{\min}:=\min_{t\le s\le T}p_s$, and set 
\[
    V_T^{\mathrm{ask}} := V_T~,
\quad
    V_T^{\mathrm{bid}} := \sum_{t=1}^{T-1}(p_t-p_t^{\min})~,
\quad\text{and}\quad
    V_T^{\mathrm{two}} := V_T^{\mathrm{ask}}+V_T^{\mathrm{bid}}
=
    \sum_{t=1}^{T-1}(\pmax_t-p_t^{\min})~.
\]
Recall that in the ask-only case primary exposure vanishes on decreasing price paths. The two-sided exposure $V_T^{\mathrm{two}}$ instead controls price movements in both directions.

Define the two-sided strategy for any sequence $\lambda_t\in[0,1]$ as $B_1=0$, $A_1=1$, and, for $t\ge2$,
\begin{equation}
\label{eq:two-sided-policy}
 B_t=m_{t-1}-(1-\lambda_t)(m_{t-1}-p_{t-1})_+~,
 \qquad
 A_t=m_{t-1}+(1-\lambda_t)(p_{t-1}-m_{t-1})_+~.
\end{equation}
The transformation $(p,m,b)\mapsto(1-p,1-m,1-b)$ turns a bid into an ask with the same execution time and reward. Its primary exposure is $V_T^{\mathrm{bid}}$, and its secondary exposure remains $W_T$. Unlike \citet{marketmakingwithoutregret}, we do not need to swap the quotes coming from the two subroutines, because these strategies are perfectly mirrored and $B_t\le A_t$ is automatically enforced.

Define the observable quantity
\begin{equation}
\label{eq:two-sided-observed}
 Z_t:=\sum_{s=1}^{t-1}
 \left(\max_{s\le u\le t}p_u-\min_{s\le u\le t}p_u\right)~,
 \qquad Z_1:=0.
\end{equation}
The sequence of $Z_t$ is nondecreasing, and $Z_T=V_T^{\mathrm{two}}$. We know show how we can achieve optimal static and dynamic regret by tuning $\lambda_t$ with this quantity. All the proofs are deferred to  Appendix~\ref{app:bid-ask-proofs}.
\begin{restatable}{theorem}{twosidedupper}
\label{thm:adaptive-two-sided-upper}
For every pair of price paths, policy~\eqref{eq:two-sided-policy} with
\begin{equation}
\label{eq:adaptive-two-sided-tuning}
 \lambda_t=\frac1t\min\left\{1,
 \frac{\sqrt{1+(H_T-1)Z_{t-1}}}{H_T-1}\right\}
 \qquad(t\ge2)
\end{equation}
satisfies
\[
 R_T^{\mathrm{two}}\le R_T^{\mathrm{dyn,two}}
 =O\!\left(1+\sqrt{V_T^{\mathrm{two}}H_T}
    +V_T^{\mathrm{two}}+W_T\right).
\]
If the horizon $T$ is unknown, the choice
\begin{equation}
\label{eq:anytime-two-sided-tuning}
 \lambda_t=\frac1t\min\left\{1,
 \frac{\sqrt{1+H_{t-1}Z_{t-1}}}{H_{t-1}}\right\}
  \qquad(t\ge2)
\end{equation}
satisfies the same bound with an additional $O(\log\log(T+1))$ term.
\end{restatable}

The proof applies the one-sided analysis twice and adds the results. Adding the two sides replaces $\pmax_t$ (reps. $\pmin_t$) by the remaining range $D_t=\pmax_t-\pmin_t$, whose total over the horizon is $V_T^{\mathrm{two}}$, because the asks are charged the decrease of $\pmax_t$ and the bids the increase of $\pmin_t$. Each round then costs the learner an amount proportional to $\lambda_t$ for the traded orders plus $(D_{t-1}-D_t)/\lambda_t$ for the orders that a moving threshold leaves behind, and balancing the two would require $V_T^{\mathrm{two}}$ in advance. The right tuning depends on how far the price will still move, which is revealed only at the end. We tune the policy using $Z_t$, which is observable and keeps track of price movements. The adversary cannot inflict a loss using price movements without enlarging $Z_t$, so the strategy becomes more cautious if price oscillates.

\begin{restatable}{theorem}{twosidedlower}
\label{thm:two-sided-lower}
For every policy, $T\ge2$, $v\in(0,T-1]$, and $w\in[0,T-1]$, there is a deterministic price path with $V_T^{\mathrm{two}}\le v$ and $W_T\le w$ such that
\[
 R_T^{\mathrm{two}}\ge
 \max\left\{\frac14,\;\frac18\sqrt{v(H_T-1)},\;
             \frac ve,\;\frac{3w}{32}\right\}
 =\Omega\!\left(1+\sqrt{v\log T}+v+w\right)~.
\]
The same lower bound holds for $R_T^{\mathrm{dyn,two}}$.
\end{restatable}

Each of the four terms comes from its own construction. The constant term and the two exposure terms are inherited from the one-sided lower bounds. The remaining term runs the construction of Theorem~\ref{thm:baseline-lower}, which hides the final execution threshold behind random price decreases, on a budget.

The two-sided bounds are tight because primary exposure accounts for both upward and downward price movements. In particular, the nonincreasing paths in Theorem~\ref{thm:baseline-lower} have zero ask exposure, but contribute to $V_T^{\mathrm{bid}}$ here. The lower bound requires $v>0$ because its secondary-exposure construction uses a positive, arbitrarily small primary-price movement. The horizon-free tuning has an additional $\log\log T$ term; the matching rate is attained by the horizon-dependent, exposure-adaptive policy.

To understand why static and dynamic regret have the same rate, consider that, on the hard nonincreasing instances of Theorem~\ref{thm:baseline-lower}, the dynamic benchmark beats the fixed one by $\sum_t (p_t - p_T)$, and since $\pmin_t \le p_T$,
\[
    \sum_{t=1}^{T-1}(p_t - p_T)
    \le \sum_{t=1}^{T-1}(p_t - \pmin_t)
    = V_T^{\mathrm{bid}}
    \le V_T^{\mathrm{two}}~.
\]
So the extra power of the dynamic comparator is bounded by a term already present in the bound. In the one-sided case that same gap was $\Theta(\sqrt T)$, because $V_T^{\mathrm{ask}}$ does not capture downward shifts.

\section{Discussion and open problems}
This work shows that market making with persistent orders is possible, even in the presence of adversarial market prices. Our analysis relies on the impossibility of canceling orders to reduce exposure to future costs. Allowing cancellations is therefore the next most natural step to make our setting more realistic. Other practical aspects that our model ignores are inventory and liquidity constraints. The extent to which violations of these constraints are allowed defines the problem as online learning with constraints (where bounds on cumulative violations are sought) or as a general state-based decision process with adversarial rewards.

\clearpage

\section*{AI Disclosure}
GPT-6.1 Sol assisted with drafting, reorganization, developing and checking proofs, reference verification, and compilation.

\section*{Acknowledgments}
NCB and LF acknowledge support by the EU Horizon CL4-2022-HUMAN-02 research and innovation action under grant agreement 101120237, project ELIAS (European Lighthouse of AI for Sustainability).
TC gratefully acknowledges the support of the Natural Sciences and Engineering Research Council of Canada (NSERC) through grant RGPIN-2023-03688 (Discovery Grants Program).

\label{page:main-end}

\bibliography{references}

\clearpage

\appendix

\section{Extension: structured price paths}
\label{app:structured}


In this extension we study how the performance of our strategies changes when we add assumptions on the price processes. Throughout, the policies remain unchanged and receive no additional information, so these improvements occur without any tuning and changes.

\subsection{Converging prices}

We start by bounding the range of oscillations of the two prices, in particular we assume that there is a pair of prices $p_\infty$ and $m_\infty$ such that $|p_t-p_\infty|\le a_t$ and $|m_t-m_\infty|\le b_t$ for all $t$, where $(a_t)_t$ and $(b_t)_t$ are nonincreasing nonnegative sequences and neither the limits nor the error bounds need to be known to the maker.

Intuitively, if the prices are converging, the exposure terms can be controlled by the rate of convergence, but we still pay the baseline term due to the fact that the two markets differ from each other. It then disappears if we assume that $m_\infty = p_\infty$.
\begin{proposition}
\label{prop:converging-prices}
For any pair $p_\infty, m_\infty \in [0, 1]$ and any pair of nonincreasing nonnegative sequences $(a_t)_t$ and $(b_t)_t$ such that $|p_t-p_\infty|\le a_t$ and $|m_t-m_\infty|\le b_t$ for all $t$.
Policy~\eqref{eq:policy} with $\lambda_t=1/t$ and $\lambda_t=1/\sqrt{t-1}$, satisfies respectively
\[
R_T \le H_T+2\sum_{t=1}^{T-1}a_t+4\sum_{t=1}^{T-1}b_t,
\qquad
R_T^{\mathrm{dyn}} \le 1+3\sqrt{T-1}+2\sum_{t=1}^{T-1}a_t+4\sum_{t=1}^{T-1}b_t.
\]
If $p_\infty=m_\infty$, either policy also satisfies
\begin{equation}
\label{eq:equalmarketregret}
R_T^{\mathrm{dyn}}\le \sum_{t=1}^T(a_t+b_t)+2\sum_{t=1}^{T-1}b_t.
\end{equation}
\end{proposition}
\begin{proof}
For every $s\ge t$, we have $p_s-p_t\le2a_t$ and $|m_s-m_t|\le2b_t$. Therefore,
\[
V_T\le2\sum_{t=1}^{T-1}a_t,\qquad W_T\le2\sum_{t=1}^{T-1}b_t,
\]
and the first two claims follow from Theorems~\ref{thm:fixed} and~\ref{thm:dynamic}.
When the limits coincide, an ask submitted at $t$ and executed at $s\ge t$ earns at most $p_s-m_s\le a_t+b_t$. This also bounds the dynamic comparator's reward for that submission date. Both policies have nonnegative reward on the first round and quote $A_t\ge m_{t-1}$ thereafter. An order submitted at $t\ge2$ consequently earns at least $-2b_{t-1}$, whether it executes or not. Summing proves the last claim.
\end{proof}
Convergence in the primary market controls $V_T$, while convergence in the secondary market controls $W_T$, so the two exposure bounds also apply separately. If both infinite price sequences converge, their tail errors vanish, so both regrets are $o(T)$, but convergence alone gives no uniform finite-horizon rate, however, since any finite hard instance can be extended with constant tails.


The baseline terms remain necessary for the unchanged policies even on constant distinct paths, so when $a_t = b_t = 0$ for all $t$ and $p_\infty \neq m_\infty$. Taking $p_t=1$ and $m_t=1-g$, with $g\in(0,1]$, gives respective regrets $g(H_T-1)$ and $g\sqrt{T-1}$.

When $p_\infty = m_\infty$, the additional dependence on the secondary error from~\eqref{eq:equalmarketregret} is necessary even for arbitrary policies. Take $p_t=1/2$ and $m_t=1/2\pm b_t$, with independent equally likely signs sampled before play, where $b_t\le1/2$ decreases to zero. Any ask has conditional expected reward at most zero, while the dynamic comparator earns $b_t$ when the cost is low and zero otherwise. Hence
\[
\E[R_T^{\mathrm{dyn}}]\ge\frac12\sum_{t=1}^T b_t.
\]
For each horizon, some deterministic realization attains this bound, and both price sequences converge to $1/2$.

\subsection{Lagged secondary prices}

We now consider the case in which the secondary market follows the primary market with a one-round delay~\citep{lo1990trading}. In particular, suppose that $m_{t+1}=p_t$ for every $t<T$, while $m_1$ is arbitrary. This assumption does not restrict the evolution of primary prices, but relates the execution cost of an order to the primary price observed in the preceding round.

This phenomenon can arise when price discovery occurs primarily in one market, while the other adjusts more slowly, for example due to differences in liquidity, trading frequency or information transmission. This creates temporary price discrepancies that the maker can exploit when outstanding orders execute, even though the two markets track the same underlying asset.

Under this assumption, an ask $a$ executing at round $s\ge 2$ earns
\[
    a-m_s=a-p_{s-1}\le p_s-p_{s-1}.
\]
Thus, positive profits can only arise from upward price movements. To capture this, define
\[
    \delta_1=(p_1-m_1)_+,
    \qquad
    \delta_t=(p_t-p_{t-1})_+
    \quad (t\ge 2),
\]
and let
\[
    J_T:=\sum_{t=1}^T\max_{t\le s\le T}\delta_s.
\]
The quantity $J_T$ measures the cumulative opportunity created by future upward price jumps. As we show next, it also controls the regret of our strategies, without requiring any change to the policy.

\begin{proposition}
Suppose that $m_{t+1}=p_t$ for all $t < T$. For any choice of $\lambda_t \in [0,1]$, policy~\eqref{eq:policy} satisfies
\[
    R_T\le R_T^{\mathrm{dyn}}
    \le J_T+\sum_{t=2}^T\lambda_t\delta_{t-1}
    \le 2J_T.
\]
\end{proposition}

\begin{proof}
Fix a submission date $t$ and an ask $a$ that executes at $s=\tau_t(a)$. Since $a\le p_s$, the lag assumption gives
\[
    r_t(a)=a-m_s\le \delta_s
    \le \max_{t\le k\le T}\delta_k.
\]
The same bound holds if the ask never executes, since its reward is zero. Summing over submission dates yields
\[
    \mathrm{OPT}_T^{\mathrm{dyn}}\le J_T.
\]

We now lower bound the reward of policy~(6). The first ask $A_1=1$ has nonnegative reward. For $t\ge 2$, suppose first that $A_t$ executes immediately. Since $m_t=p_{t-1}$, the policy definition gives
\[
    r_t(A_t)=A_t-p_{t-1}
    \ge -\lambda_t(p_{t-1}-m_{t-1})_+
    =-\lambda_t\delta_{t-1}.
\]
If instead the ask executes at some $s>t$, it must have remained unexecuted at round $s-1$, so $A_t>p_{s-1}=m_s$. Consequently, its reward is positive. An ask that never executes has zero reward. Therefore, in every case,
\[
    r_t(A_t)\ge-\lambda_t\delta_{t-1}.
\]
Summing over rounds, we obtain
\[
    \sum_{t=1}^T r_t(A_t)
    \ge-\sum_{t=2}^T\lambda_t\delta_{t-1}.
\]
Combining this with the bound on the dynamic comparator proves the first inequality. Finally, since $\lambda_t\le 1$,
\[
    \sum_{t=2}^T\lambda_t\delta_{t-1}
    \le\sum_{t=1}^{T-1}\delta_t
    \le J_T,
\]
which concludes the proof.
\end{proof}

The new bound can be substantially smaller than the general exposure bound. For example, consider the steadily increasing price paths
\[
    p_t=\frac{t}{T},
    \qquad
    m_t=\frac{t-1}{T}.
\]
These satisfy the lag assumption and give
\[
    J_T=1,
    \qquad
    V_T=W_T=\frac{T-1}{2}.
\]
Our general guarantees therefore scale linearly with $T$, whereas the proposition gives constant regret for both comparators. More generally, if $\delta_t\le c/t$ for every $t$, then $J_T\le cH_T$, and both regrets are $O(\log T)$.

The lag assumption alone, however, does not guarantee sublinear regret. Indeed, the primary-exposure lower-bound construction of Section~\ref{sec:lower} already satisfies $m_{t+1}=p_t$: primary prices remain zero until the final round, while secondary prices are identically zero. In that construction, a single upward price jump creates $J_T=\Theta(T)$, and every policy can incur linear regret. Hence, controlling the magnitude and timing of upward price movements is still necessary.

\clearpage

\section{Proof of the prefix comparison}
\label{app:potential}

%

\begin{proofof}{Lemma~\ref{lem:potential}}
Recall that $\pmax_t=\max_{t\le s\le T}p_s$ is the highest remaining primary price, $C_t=1+\sum_{s=1}^{t-1}m_s$, and $H_t=\sum_{s=1}^t1/s$. The quantity $Q_t^\circ$ sums the eventual surrogate rewards of orders submitted on the first $t$ rounds; those orders need not have executed by round $t$. The expression $t\pmax_t-C_t$ bounds the surrogate profit of a fixed quote whose orders execute for precisely the first $t$ submission dates, as shown in~\eqref{eq:prefix}.

We prove the first inequality by induction. At $t=1$, the auxiliary quote is $1$, its surrogate reward is zero, and $C_1=H_1=1$. Thus the required inequality is $0\ge\pmax_1-1$, which follows from $\pmax_1\le1$.

Fix $t\ge2$ and assume the inequality holds at $t-1$. We distinguish three cases according to the new order's surrogate reward.
\begin{itemize}
\item\textbf{A positive-margin order executes.} Suppose $\pmax_{t-1}>m_{t-1}$ and $\surrA_t\le\pmax_t$. Since $\lambda_t=1/t$,
\[
 r_t'(\surrA_t)=(1-\lambda_t)(\pmax_{t-1}-m_{t-1})
 =\frac{t-1}{t}(\pmax_{t-1}-m_{t-1}).
\]
Add this reward to the induction hypothesis and use $C_t=C_{t-1}+m_{t-1}$ and $H_t=H_{t-1}+1/t$. Subtracting the desired lower bound gives
\begin{align*}
 Q_t^\circ-(t\pmax_t-C_t-H_t+1)
 &\ge\left(t-\frac1t\right)(\pmax_{t-1}-\pmax_t)
       +\frac{1+m_{t-1}-\pmax_t}{t}\ge0.
\end{align*}
Both terms are nonnegative: the remaining primary maximum cannot increase, and all prices lie in $[0,1]$.
\item\textbf{The reference margin is nonpositive.} If $\pmax_{t-1}\le m_{t-1}$, the quote is $m_{t-1}$ and its surrogate reward is zero. Also,
\[
 t\pmax_t\le t\pmax_{t-1}\le(t-1)\pmax_{t-1}+m_{t-1}.
\]
\item\textbf{A positive-margin order does not execute.} If $\pmax_{t-1}>m_{t-1}$ and $\surrA_t>\pmax_t$, its reward is again zero. The policy definition gives
\[
 t\pmax_t<t\surrA_t=(t-1)\pmax_{t-1}+m_{t-1}.
\]
\end{itemize}
In each zero-reward case, subtracting $C_t=C_{t-1}+m_{t-1}$ shows that
\[
 t\pmax_t-C_t\le(t-1)\pmax_{t-1}-C_{t-1}.
\]
Thus the quantity we need to match has not increased, whereas $Q_t^\circ=Q_{t-1}^\circ$ and $H_t\ge H_{t-1}$. The induction hypothesis therefore gives the required inequality at $t$.

To compare with the best fixed quote, we must cover every possible prefix length. All auxiliary surrogate rewards are nonnegative, so, for every $k\le T$,
\[
 Q_T^\circ\ge Q_k^\circ\ge k\pmax_k-C_k-H_k+1
 \ge k\pmax_k-C_k-H_T+1.
\]
Together with $Q_T^\circ\ge0$, this implies
\[
 \max\left\{0,\max_{1\le k\le T}(k\pmax_k-C_k)\right\}
 \le Q_T^\circ+H_T-1.
\]
The benchmark bound~\eqref{eq:prefix} now yields $R_T'(\surrA)=\OPT_T'-Q_T^\circ\le H_T-1$.
\end{proofof}

\section{The static-regret tuning can incur linear dynamic regret}
\label{app:separation}

The static and dynamic regret bounds use the same policy~\eqref{eq:policy}, with different choices of $\lambda_t$. We show that the static choice $\lambda_t=1/t$ can have dynamic regret at least $T/12$, even when $V_T=W_T=0$. The construction keeps prices constant for the first half of the horizon, then makes them fall slightly more than the amount subtracted by the static policy from its quote.

Let $T=2n$ with $n\ge2$, set $m_t=0$ on every round, and choose
\[
 p_t=1\quad(t\le n),\qquad
 p_t=\left(1-\frac2t\right)p_{t-1}\quad(t>n).
\]
The primary prices are positive and nonincreasing, so $V_T=W_T=0$. For every $t>n$, the static choice gives
\[
 A_t=\left(1-\frac1t\right)p_{t-1}
     >\left(1-\frac2t\right)p_{t-1}=p_t.
\]
This order does not execute on its submission round. Since primary prices never increase, it cannot execute later either. However, the available sale price remains bounded away from zero:
\[
 p_t=\prod_{s=n+1}^{t}\frac{s-2}{s}
     =\frac{n(n-1)}{t(t-1)}\ge\frac16
     \qquad(n<t\le2n).
\]
The dynamic comparator quotes $p_t$ and earns $p_t$ immediately on each of these $n$ rounds. Regret on every other round is nonnegative, so $R_T^{\mathrm{dyn}}\ge n/6=T/12$.

\section{Details for the lower bounds}
\label{app:lower}

\subsection{The last primary price need not be the best fixed quote}
\label{app:terminal-quote}

The static-regret part of Theorem~\ref{thm:baseline-lower} uses the reward of the fixed quote $p_T$ as a lower bound on $\OPT_T$. That quote need not be optimal. To see this, consider any nonincreasing primary path and a constant secondary price $c<p_T$. An order either executes immediately or never executes. Hence a fixed quote $a$ trades on a prefix of submission dates. If that prefix has length $k\ge1$, its total reward is
\[
 k(a-c)\le k(p_k-c).
\]
Conversely, quoting $p_k$ trades on at least $k$ rounds, each with positive margin. A quote that never trades earns zero, so
\[
 \OPT_T=\max_{1\le k\le T}k(p_k-c).
\]
For example, with $p=(1,1/4)$ and $m=(0,0)$, quote $1$ earns $1$, while quote $p_T=1/4$ earns only $1/2$.

\paragraph{Averaging over price sequences.}
In the next two proofs, we fix an arbitrary policy and choose the entire price sequence randomly, independently of the policy's internal randomness. We bound the average of the policy's expected regret over these sequences. At least one fixed sequence must attain that average, which proves the desired deterministic lower bound.

\subsection{Primary exposure}
\label{app:primary-lower}


\begin{proofof}{Theorem~\ref{thm:primary-lower}}
The construction hides the only profitable primary price until the final round, after every quote has been chosen. Let $\beta=v/(T-1)\in(0,1]$, set all secondary prices to zero, and take $p_t=0$ for $t<T$ and $p_T=X$. Choose $X\in[\beta/e,\beta]$ so that
\[
 \mathbb P(X\ge a)=
 \begin{cases}
 1,&0\le a\le\beta/e,\\
 \beta/(ea),&\beta/e<a\le\beta,\\
 0,&a>\beta.
 \end{cases}
\]
This distribution assigns probability $1/e$ to $X=\beta$, so the displayed probability includes execution at the endpoint. Every positive ask $a$ executes only at $T$, and does so exactly when $X\ge a$; a zero ask earns zero. The distribution is chosen to make the expected reward of every ask at most $\beta/e$:
\[
 a\mathbb P(X\ge a)\le\beta/e.
\]
The prices observed before the last round provide no information about $X$. Thus this bound also applies to each quote chosen by a randomized policy, including its final quote. The learner's expected total reward is at most $T\beta/e$.

The fixed comparator knows $X$ and quotes it on every round, earning $TX$. Moreover,
\[
 \E[X]=\int_0^\beta\mathbb P(X\ge x)\,dx
       =\frac\beta e+\int_{\beta/e}^{\beta}\frac{\beta}{ex}\,dx
       =\frac{2\beta}{e}.
\]
The average regret is therefore at least $T\beta/e=Tv/(e(T-1))\ge v/e$. Every realization satisfies $V_T=(T-1)X\le v$ and $W_T=0$. Selecting a value of $X$ with at least the average expected regret proves the claim.
\end{proofof}

\subsection{Secondary exposure}
\label{app:secondary-lower}


\begin{proofof}{Theorem~\ref{thm:secondary-lower}}
If $w=0$, take $p_t=m_t=0$ on every round. Suppose $w>0$, and choose
\[
 \delta=\frac{w}{T-1},\qquad 0<h\le\frac{\delta}{4T}.
\]
Before play begins, choose one of two scenarios, each with probability $1/2$. In both scenarios, the primary price is $3/4$ and the secondary price is $1/2$ on every round before $T$. On the last round, the primary price increases by $h$. The only difference between the scenarios is the final secondary price: it falls by $\delta$ in one and rises by $\delta$ in the other. Each row of the table below specifies both market prices for one scenario.
\[
\begin{array}{l|cc|cc}
 &\multicolumn{2}{c|}{\text{Rounds }t<T}
 &\multicolumn{2}{c}{\text{Round }T}\\
\text{Scenario}&\text{Primary}&\text{Secondary}&\text{Primary}&\text{Secondary}\\\hline
\text{Secondary price falls}&3/4&1/2&3/4+h&1/2-\delta\\
\text{Secondary price rises}&3/4&1/2&3/4+h&1/2+\delta
\end{array}
\]
All prices lie in $[0,1]$, since $\delta\le1/4$ and $h\le\delta/(4T)$. The maker observes the same prices in both scenarios before round $T$. Its final quote is also chosen before observing that round's prices. Thus, when submitting any of its asks, the maker has no information about which scenario was chosen.

For $t<T$, an ask $a\le3/4$ executes immediately and earns $a-1/2$. An ask $3/4<a\le3/4+h$ waits until round $T$, when the primary price first reaches it; such asks exist because $h>0$. Its hedging cost is then $1/2-\delta$ or $1/2+\delta$, each with probability $1/2$, so its expected reward is again $a-1/2$. Asks above $3/4+h$ never execute. An ask submitted at $T$ also has expected reward $a-1/2$ if $a\le3/4+h$, and zero otherwise. Therefore, each quote has expected reward at most $1/4+h$, and the learner's expected total reward is at most $T(1/4+h)$.

When costs fall, the fixed quote $3/4+h$ delays every order until $T$ and earns $T(1/4+h+\delta)$. When costs rise, the fixed quote $3/4$ trades immediately on every round and earns $T/4-\delta$; only its last order pays the higher cost. These are valid choices for the fixed comparator in the respective sequences. Averaging their rewards and subtracting the learner's upper bound gives regret at least
\[
 \frac{T-1}{2}\delta-\frac T2h
 \ge\frac w2-\frac\delta8\ge\frac38w,
\]
since $\delta=w/(T-1)\le w$.

In both scenarios, $w_t=\delta$ for every $t<T$, so $W_T=(T-1)\delta=w$, and the primary exposure is $V_T=(T-1)h$. At least one of the two deterministic price paths has regret at least the average and therefore satisfies
\[
 W_T=w,\qquad V_T=(T-1)h,\qquad R_T\ge\frac38w.
\]
Taking $h=\delta/(4T)$ gives $V_T=w/(4T)$, proving the first claim. By choosing $h>0$ arbitrarily small, we obtain deterministic price paths with arbitrarily small $V_T$ while retaining $W_T=w$ and $R_T\ge3w/8$, proving the final claim.
\end{proofof}

\section{Details of the simultaneous bids and asks proofs}
\label{app:bid-ask-proofs}

\begin{proofof}{Theorem~\ref{thm:adaptive-two-sided-upper}}
The proof follows Theorem~\ref{thm:dynamic}. Write $D_t:=\pmax_t-p_t^{\min}$, so that $D_T=0$ and $\sum_{t=1}^{T-1}D_t=V_T^{\mathrm{two}}$.
%
The learner's reward splits over the two sides by definition, so the proof of Lemma~\ref{lem:execution-cost} contributes $1+2W_T$ on each and Lemma~\ref{lem:fixed-surrogate} contributes $V_T^{\mathrm{ask}}$ and $V_T^{\mathrm{bid}}$ respectively. Bounding what remains by~\eqref{eq:combining} on each side and adding the two gives, for any $\lambda_t\in(0,1]$,
\begin{align}
      R_T^{\mathrm{dyn,two}}
  &\le
      2+4W_T+V_T^{\mathrm{ask}}+V_T^{\mathrm{bid}}
      +\sum_{t=2}^T\left(2\lambda_t+\frac{(\pmax_{t-1}-\pmax_t)+(\pmin_t-\pmin_{t-1})}{\lambda_t}\right)
  \nonumber\\
  &\le
      2+V_T^{\mathrm{two}}+4W_T+2\sum_{t=2}^T\lambda_t+\sum_{t=2}^T\min\left\{2,\frac{D_{t-1}-D_t}{\lambda_t}\right\}~.
  \label{eq:two-sided-master}
  \end{align}
The last term is truncated by $2$ because the bid and ask side each contributes to the per-round surrogate regret $\max_a r_t'(a)-r_t'(\surrA_t)$ of the clairvoyant policy, which is at most $1$ because the optimal surrogate reward is at most $1$ while the clairvoyant quote never crosses $m_{t-1}$ and therefore has nonnegative surrogate reward. Summing the two sides bounds the contribution of round $t$ by $\min\{2,2\lambda_t+(D_{t-1}-D_t)/\lambda_t\}$, and since $\lambda_t\ge0$ the truncation at $2$ may be carried by the second summand alone.

Suppose first that $V_T^{\mathrm{two}}>0$ is known. Take $\lambda_t=\alpha/t$, where
$\alpha=\min\{1,\sqrt{V_T^{\mathrm{two}}/(H_T-1)}\}$.
Since
\begin{equation}
\label{eq:telescoping-D}    
 \sum_{t=2}^T t(D_{t-1}-D_t)
 =D_1+\sum_{t=1}^{T-1}D_t\le2V_T^{\mathrm{two}},
\end{equation}
\eqref{eq:two-sided-master} yields
\[
 R_T^{\mathrm{dyn,two}}
 \le2+2\alpha(H_T-1)+\frac{2V_T^{\mathrm{two}}}{\alpha}
  +V_T^{\mathrm{two}}+4W_T~,
\]
which gives the claimed rate. If $V_T^{\mathrm{two}}=0$, all primary prices are constant and taking $\lambda_t=0$ gives the corresponding constant baseline bound.

We now remove knowledge of $V_T^{\mathrm{two}}$. The relevant property of $Z_t$ is that, for every $2\le s\le T$,
\begin{equation}
\label{eq:two-sided-prefix}
 \sum_{t=2}^s t(D_{t-1}-D_t)
 =(D_1-D_s)+\sum_{j=1}^{s-1}(D_j-D_s)
 \le2Z_s~,
\end{equation}
which mirrors~\eqref{eq:telescoping-D}. To see that it holds, note that, for $1 \le j < s \le T$,
\[
    D_j - D_s = (\pmax_j - \pmax_s) - (\pmin_j - \pmin_s) \le \max_{j \le u \le s} p_u - \min_{j \le u \le s} p_u~.
\]
%
For the tuning~\eqref{eq:adaptive-two-sided-tuning}, let $\lambda_t = \alpha_t/t$. Since $Z_t$ is nondecreasing, $\alpha_t$ is also nondecreasing, and
\[
    \sum_{t=2}^T\lambda_t
\le
    (H_T-1)\alpha_T
\le
    \sqrt{1+(H_T-1)V_T^{\mathrm{two}}}~.
\]
To bound the last sum in \eqref{eq:two-sided-master}, group rounds by
$2^k\le1+(H_T-1)Z_{t-1}<2^{k+1}$. Within group $k$,
$\alpha_t\ge\min\{1,2^{k/2}/(H_T-1)\}$.
If $b$ is the last round in group $k$, \eqref{eq:two-sided-prefix} bounds the contribution of all earlier rounds by
\[
 \frac{\sum_{t=2}^{b-1}t(D_{t-1}-D_t)}{\min\{1,2^{k/2}/(H_T-1)\}}
 \le4\max\left\{2^{k/2},\frac{2^k}{H_T-1}\right\}~.
\]
The final round contributes at most $2$ by truncation. Summing over the geometric groups gives
\[
 \sum_{t=2}^T\min\left\{2,\frac{D_{t-1}-D_t}{\lambda_t}\right\}
 =O\!\left(1+\sqrt{(H_T-1)V_T^{\mathrm{two}}}
      +V_T^{\mathrm{two}}\right)~.
\]
Substitution into \eqref{eq:two-sided-master} proves the horizon-dependent guarantee.

The previous tuning, while being implementable, requires knowledge of the time horizon.


The tuning~\eqref{eq:anytime-two-sided-tuning} differs from~\eqref{eq:adaptive-two-sided-tuning} only in replacing the constant $H_T-1$ by $H_{t-1}$, we show that this costs at most a factor of two in $\alpha_t$. Dropping either term under the square root gives $\sqrt{1+H_{t-1}Z_{t-1}}\ge\max\{1,\sqrt{H_{t-1}Z_{t-1}}\}$, and dividing by $H_{t-1}\le H_T$ gives
\[
    \frac{\sqrt{1+H_{t-1}Z_{t-1}}}{H_{t-1}}
\ge
    \max\left\{\frac{1}{H_T},\sqrt{\frac{Z_{t-1}}{H_T}}\right\}
\ge
    \frac12\cdot\frac{\sqrt{1+H_TZ_{t-1}}}{H_T}~,
\]
where the last step uses $\sqrt{1+H_TZ_{t-1}}\le1+\sqrt{H_TZ_{t-1}}$. Hence $\alpha_t\ge\frac12\min\{1,\sqrt{1+H_TZ_{t-1}}/H_T\}$.

Grouping the rounds by $2^k\le1+H_TZ_{t-1}<2^{k+1}$ now repeats the argument above with $H_T$ in place of $H_T-1$, the factor $\frac12$ only doubling the constant obtained for each group, so the last sum in~\eqref{eq:two-sided-master} is again $O(1+\sqrt{H_TV_T^{\mathrm{two}}}+V_T^{\mathrm{two}})$.

What remains is $\sum_{t=2}^T\lambda_t$, which is where the extra term appears. Using $Z_{t-1}\le V_T^{\mathrm{two}}$ and the same square-root inequality,
\[
    \lambda_t
\le
    \frac{1+\sqrt{H_{t-1}V_T^{\mathrm{two}}}}{tH_{t-1}}
=
    \frac{1}{tH_{t-1}}+\frac{\sqrt{V_T^{\mathrm{two}}}}{t\sqrt{H_{t-1}}}~.
\]
Since $H_{t-1}\ge\log t$, comparing the two sums with the integrals of $1/(x\ln x)$ and $1/(x\sqrt{\ln x})$ gives
\[
    \sum_{t=2}^T\lambda_t
=
    O\!\left(\log H_T+\sqrt{H_TV_T^{\mathrm{two}}}\right)~.
\]
The second term already appears in the horizon-dependent bound. This concludes the proof.
\end{proofof}

\begin{proofof}{Theorem~\ref{thm:two-sided-lower}}
As in Section~\ref{sec:lower}, we construct random oblivious price paths and select a deterministic realization with at least the average regret. For the constant term, take $m_t=0$ and $p_t=X$ on every round, where $X$ is equally likely to be $1/2$ or $1$. The fixed comparator earns $TX$, while the learner's expected reward on the first round is at most $1/2$, giving regret at least $1/4$ with both exposures zero. Theorems~\ref{thm:primary-lower} and~\ref{thm:secondary-lower} give the $\Omega(v)$ and $\Omega(w)$ terms. Their constructions satisfy $p_t\ge m_t$, so bids cannot earn positive reward. For the secondary construction, its positive primary increment can be chosen arbitrarily small to respect any $v>0$; for $w>(T-1)/4$, apply Theorem~\ref{thm:secondary-lower} with budget $(T-1)/4=\Omega(w)$.

It remains to obtain $\Omega(\sqrt{v(H_T-1)})$, and we may assume $1/(H_T-1)\le v\le H_T-1$, since outside this range the constant term or the term $v/e$ is already of the required order. The construction of Theorem~\ref{thm:baseline-lower} hides the final execution threshold behind random price decreases, but here each decrease also costs primary exposure, so its size must be tuned against the budget. If the price falls by a factor $\delta$ with probability $\delta$, the regret of that round is proportional to $\delta$, because the maker must quote below the current price to avoid missing execution, while its expected exposure is proportional to $\delta^2$, because a decrease costs only when it occurs and its size is again $\delta$. Regret is therefore linear and exposure quadratic in the same parameter, so shrinking it buys budget quadratically while costing regret only linearly. Taking $\delta$ proportional to $1/t$ makes the two sums $\eta(H_T-1)$ and $\eta^2(H_T-1)$, so the budget $v$ fixes $\eta$ at the order of $\sqrt{v/(H_T-1)}$ and leaves regret of order $\sqrt{v(H_T-1)}$.

Following Theorem~\ref{thm:baseline-lower}, set $p_1=1$, $m_t=0$ and $\eta:=\frac14\sqrt{v/(H_T-1)}$, so that $v\le H_T-1$ gives $\eta\le1/4$. At each $t\ge2$, let $p_t=(1-\eta/t)p_{t-1}$ with probability $\eta/t$ and $p_t=p_{t-1}$ otherwise, independently of the previous draws, until $\lfloor v/\eta\rfloor$ decreases have occurred, after which the price stays constant. This cap is what makes the budget hold on every realization rather than only in expectation, and it is loose enough to bind rarely. A decrease at round $t$ raises the exposure by $(t-1)(\eta/t)p_{t-1}\le\eta$, so
\[
    V_T^{\mathrm{two}}=V_T^{\mathrm{bid}}
    =\sum_{t=2}^T(t-1)(p_{t-1}-p_t)
    \le\left\lfloor\frac v\eta\right\rfloor\eta\le v,
    \qquad W_T=0~.
    \qquad W_T=0~.
\]
Every path is nonincreasing, so $V_T^{\mathrm{ask}}=0$, and a bid earns $-b$ when it executes, so the best bid is $0$ and the two-sided benchmark coincides with the ask-only one.

Let $I_t$ indicate that the cap has not yet bound at round $t$. Conditioning on the first $t-1$ rounds, the argument of Theorem~\ref{thm:baseline-lower} bounds the expected reward of any ask by $(1-I_t\eta/t)p_{t-1}$, while the price itself satisfies $\E[p_t\mid p_{t-1}]=(1-I_t\eta^2/t^2)p_{t-1}$, because a decrease of relative size $\eta/t$ occurs with probability $\eta/t$. The fixed ask $p_T$ earns $Tp_T$, so the telescoping calculation of that theorem gives
\[
    \E[R_T^{\mathrm{two}}]
    \ge\sum_{t=2}^T\E\!\left[I_tp_{t-1}\left(\frac{\eta}{t}-t\,\frac{\eta^2}{t^2}\right)\right]
    =\eta(1-\eta)\sum_{t=2}^T\frac{\E[I_tp_{t-1}]}{t}~.
\]
This is where the harmonic sum appears, and it remains to bound $\E[I_tp_{t-1}]$ below by a constant independent of $t$.

Two events can make $\E[I_tp_{t-1}]$ small, the cap having already bound and the price having already fallen. We rule out both uniformly in $t$. The expected number of decreases is at most $\sum_{t=2}^T\eta/t=\eta(H_T-1)$, and $v/\eta=4\sqrt{v(H_T-1)}\ge4$ by the assumption $v\ge1/(H_T-1)$, so $\lfloor v/\eta\rfloor\ge v/(2\eta)$ and Markov's inequality gives $\mathbb P(I_t=0)\le2\eta^2(H_T-1)/v=1/8$. Iterating $\E[p_t]\ge(1-\eta^2/t^2)\E[p_{t-1}]$ gives $\E[p_{t-1}]\ge1-\eta^2\sum_{s\ge2}s^{-2}\ge15/16$. Since $p_{t-1}\le1$, the two estimates combine into $\E[I_tp_{t-1}]\ge13/16$, and $\eta\le1/4$ gives $1-\eta\ge3/4$, so
\[
    \E[R_T^{\mathrm{two}}]
    \ge\frac14\sqrt{\frac{v}{H_T-1}}\cdot\frac34\cdot\frac{13}{16}(H_T-1)
    =\frac{39}{256}\sqrt{v(H_T-1)}
    \ge\frac18\sqrt{v(H_T-1)}~.
\]
Every realization respects both budgets, so some deterministic price path attains this bound. Taking the maximum of the four constructions proves the claim, since the maximum is within a constant factor of their sum, and $R_T^{\mathrm{dyn,two}}\ge R_T^{\mathrm{two}}$ extends it to the dynamic comparator.
\end{proofof}

\end{document}